\documentclass[showpacs,preprintnumbers,amsmath,twocolumn,amssymb]{revtex4}
\usepackage{graphicx}
\usepackage{dcolumn}
\usepackage{bm}
\usepackage{amssymb}
\usepackage{array}
\usepackage{hhline}
\usepackage{longtable}

\begin{document}

\title{Standard random walks and trapping on the Koch network \\ with scale-free behavior and small-world effect}

\author{Zhongzhi Zhang $^{1,2}$}
\email{zhangzz@fudan.edu.cn}

\author{Shuigeng Zhou$^{1,2}$}
\email{sgzhou@fudan.edu.cn}

\author{Wenlei Xie $^{1,2}$}

\author{Lichao Chen $^{1,2}$}

\author{Yuan Lin$^{1,2}$}

\author{Jihong Guan$^{3}$}
\email{jhguan@tongji.edu.cn}

\affiliation {$^{1}$School of Computer Science, Fudan University,
Shanghai 200433, China}

\affiliation {$^{2}$Shanghai Key Laboratory of Intelligent
Information Processing, Fudan University, Shanghai
200433, China}

\affiliation{$^{3}$Department of Computer Science and Technology,
Tongji University, 4800 Cao'an Road, Shanghai 201804, China}

\begin{abstract}
A vast variety of real-life networks display the ubiquitous presence
of scale-free phenomenon and small-world effect, both of which play
a significant role in the dynamical processes running on networks.
Although various dynamical processes have been investigated in
scale-free small-world networks, analytical research about random
walks on such networks is much less. In this paper, we will study
analytically the scaling of the mean first-passage time (MFPT) for
random walks on scale-free small-world networks. To this end, we
first map the classical Koch fractal to a network, called Koch
network. According to this proposed mapping, we present an iterative
algorithm for generating the Koch network, based on which we derive
closed-form expressions for the relevant topological features, such
as degree distribution, clustering coefficient, average path length,
and degree correlations. The obtained solutions show that the Koch
network exhibits scale-free behavior and small-world effect. Then,
we investigate the standard random walks and trapping issue on the
Koch network. Through the recurrence relations derived from the
structure of the Koch network, we obtain the exact scaling for the
MFPT. We show that in the infinite network order limit, the MFPT
grows linearly with the number of all nodes in the network. The
obtained analytical results are corroborated by direct extensive
numerical calculations. In addition, we also determine the scaling
efficiency exponents characterizing random walks on the Koch
network.
\end{abstract}


\date{\today}
\maketitle

\section{Introduction}

Complex networks have been acknowledged as an invaluable tool for
describing real-world systems in nature and
society~\cite{AlBa02,DoMe02,Ne03,BoLaMoChHw06,DoGoMe08}. Extensive
empirical studies have uncovered that a lot of real networks share
several remarkable features~\cite{CoRoTrVi07}. One of the most
relevant is the scale-free behavior, that is, various real networks
exhibit a power-law degree distribution $P(k) \sim
k^{-\gamma}$~\cite{BaAl99}. Another very important observation is
that most real-life systems are characterized by ubiquitous
small-world effect~\cite{WaSt98}, including large clustering
coefficient~\cite{ZhZh07} and small average path length
(APL)~\cite{ZhZhChFaZhGu09}. Both the scale-free behavior and the
small-world effect have a profound impact on almost all dynamical
processes taking place on the
networks~\cite{Ne03,BoLaMoChHw06,DoGoMe08,AlJeBa00,CaNeStWa00,CoErAvHa00,CoErAvHa01,PaVe01a,WaSt98,BaWe00,ZhZhZoCh08}.

Among various dynamical processes, random walks on networks are
fundamental to many branches of science, and have received
considerable attention from the scientific
community~\cite{NoRi04,SoRebe05,CoBeTeVoKl07,BeMeTeVo08,CoTeVoBeKl08,GaSoHaMa07,BaCaPa08,LeYoKi08}.
As a primary dynamical process, random walks are related to a
plethora of other dynamics such as transport in media~\cite{HaBe87},
disease spreading~\cite{LlMa01}, and target
search~\cite{JaBl01,Sh05} to name a few. On the other hand, random
walks are useful for the study of topological structure (e.g.
betweenness and average path length~\cite{NoRi04,LeYoKi08}) and
community detection~\cite{NeGi04} on networks. In particular, as an
integral theme of random walks, trapping is related to a wide
variety of contexts~\cite{CaAb08}, such as photon-harvesting
processes in photosynthetic cells~\cite{Mo69,WhGo99} and
characterizing similarities between the elements of a
database~\cite{FoPiReSa07}. It is thus of theoretical and practical
interests to study standard random walks and trapping problem on
complex networks.

The main interesting quantity closely related to random walks is the
mean first-passage time (MFPT), and a central issue in the study of
random walks is how the MFPT scales with the size of the
system~\cite{HaBe87,MeKl04,BuCa05}. In the past several years, a lot
of endeavors have been devoted to studying the intrinsic relations
between the scaling behavior of random walks and the underlying
topological structure of
networks~\cite{NoRi04,SoRebe05,CaAb08,HuXuWuWa06}. The results of
these investigations uncovered many unusual and exotic features of
complex networks, especially of small-world and scale-free networks.
In spite of their useful insight, most of previous jobs focused on
numerical investigation, the analytical results of MFPT (in
particular the average for all pairs of nodes) for standard random
walks and trapping problem have been far less reported, with the
exception of some graphs with simple topology, such as regular
lattices~\cite{Mo69}, Sierpinski
fractals~\cite{KaBa02PRE,KaBa02IJBC}, \emph{T} fractal~\cite{Ag08},
and deterministic scale-free trees~\cite{Bobe05,ZhZhZhYiGu09}, as
well as other
structures~\cite{CoBeTeVoKl07,BeMeTeVo08,CoTeVoBeKl08}, and
exhaustive analytical research on scale-free and small-world
networks with loops is still missing.

In this paper, we analytically investigate the scaling behavior of
MFPT on complex networks with scale-free phenomenon and small-world
effect. To achieve this goal, we first propose a relevant
deterministic network, named Koch network, which is based on the
classical fractal---Koch curve. We then suggest a minimal iterative
algorithm generating the Koch network, on the basis of which we give
in detail a scrutiny of the network architecture. The analysis
results show that the Koch network is simultaneously scale free,
small world, and has large clustering coefficient. Particularly, we
show that the Koch network is completely two-point uncorrelated by
exactly computing two relevant quantities of two-point correlations,
which has never been previously found in other deterministically
growing networks. Eventually, we study the scalings of MFPT on the
proposed network for the standard random walks, together with a
special case of trapping problem with the trap fixed at a hub node.
We present that the MFPT averaged over all couples of nodes scales
linearly with the network order (total number of nodes). We derive a
closed-form solution for the MFPT characterizing the trapping
process, which also grows as a linear function of the network order.
We also compare the our results with those previously obtained for
other scale-free networks.

\begin{figure}
\begin{center}
\includegraphics[width=.75\linewidth,trim=100 20 100 0]{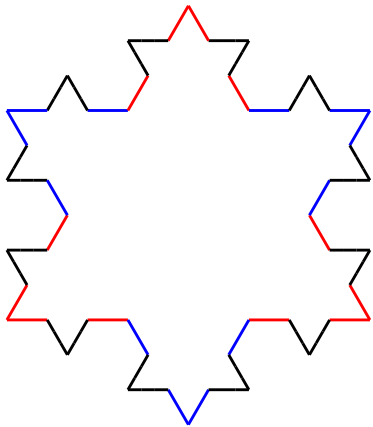} \
\end{center}
\caption[kurzform]{\label{curve} (Color online) The first two
generations of the construction for Koch curve.}
\end{figure}

\section{Network construction}

The network under consideration is derived from the Koch curve. To
define the network, we first introduce the classical fractal, Koch
island, also known as Koch curve, Koch snowflake or Koch triangle,
which was proposed by Koch~\cite{Ko1906}. This well-known fractal
denoted by $S_t$ after $t$ generations is constructed as
follows~\cite {LaVaMeVa87}. Start with an equilateral triangle and
denote this initial configuration as $S_0$. Perform a trisection of
each side of this initial triangle and construct an equilateral
triangle on each middle segment, so that the interior of the added
triangle lies in the exterior of the base triangle, then remove the
segment upon which the new triangle is established. Thus, we get
$S_1$. For each line segment in $S_1$, trisect it and draw an
equilateral triangle based on the resultant middle small segment to
obtain $S_2$. Repeat recursively the procedure of trisection of
existing line segments in last generation and addition of triangles.
In the infinite $t$ limit, we obtain the famous Koch curve $S_t$,
whose Hausdorff dimension is $d_f=\frac{2\ln 2}{\ln3}$~\cite{Ma82}.
In Fig.~\ref{curve}, we show schematically the structure of $S_{2}$.
In fact, this fractal can be easily generalized to other
dimensions~\cite{LaVaMeVa87}.

\begin{figure}
\begin{center}
\includegraphics[width=1.0\linewidth,trim=80 25 80 10]{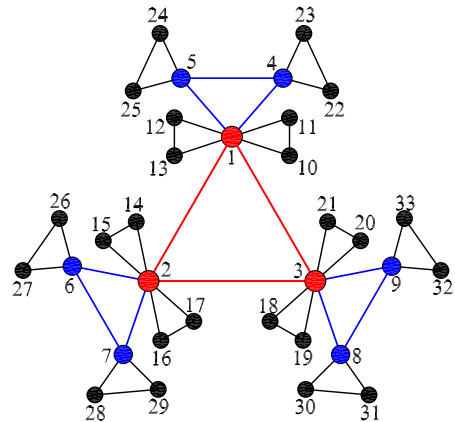}
\end{center}
\caption[kurzform]{\label{network} (Color online) Construction for
the Koch network and labels of its nodes, showing the first two
steps.}
\end{figure}

From the Koch curve we can easily construct a network (called Koch
network) using a simple mapping as follows. In the Koch network,
nodes (vertices and sites) correspond to the sides (excluding those
deleted) of the triangles constructed at all generations of the Koch
curve as shown in Fig.~\ref{curve}. That is to say, for every
triangle created at some generation, its two newly-born sides are
mapped to two nodes, while the removed side is not. We make two
nodes connected if the corresponding two sides of the Koch curves
contact each other. For uniformity, the three sides of the initial
equilateral triangle of $S_0$ also correspond to three different
nodes. Note that after the birth of each side of a triangle
constructed at a given generation, although some segments of it will
be deleted at subsequent steps, we look upon its remaining segments
as a whole and map it to only one node. Figure~\ref{network} shows
the network associated with $S_2$.

\section{Generation algorithm of Koch network}

\begin{figure}
\begin{center}
\includegraphics[width=.4\linewidth,trim=100 40 100 0]{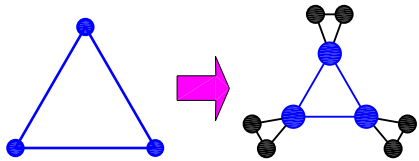}
\end{center}
\caption[kurzform]{\label{iterative} (Color online) Iterative
construction method for the network. }
\end{figure}

According to the construction process of Koch curve and the proposed
mapping from Koch curve to Koch network, we can introduce with ease
an iterative algorithm to create Koch network, denoted by $K(t)$
after $t$ generation evolutions. The algorithm is as follows.
Initially ($t=0$), $K(0)$ consists of three nodes forming a
triangle. Then, each of the three nodes of the initial triangle
gives birth to two nodes. These two new nodes and its mother node
are linked to each other shaping a new triangle. Thus we get $K(1)$
(see Fig.~\ref{iterative}). For $t\geq 1$, $K(t)$ is obtained from
$K(t-1)$. We replace each of the existing triangles of $K(t-1)$ with
the connected cluster on the right-hand side (rhs) of
Fig.~\ref{iterative} to obtain $K(t)$. The growing process is
repeated until the network reaches a desired order.
Figure~\ref{network} shows the network growth process for the first
two steps.

Next we compute the order and the size (number of all edges) of Koch
network $K(t)$. To this end, we first calculate the total number of
triangles existing at step $t$, which we denote as $L_\Delta(t)$. By
construction, this quantity increases by a factor of 4, i.e.,
$L_\Delta(t)=4\,L_\Delta(t-1)$. Considering the initial condition
$L_\Delta(0)=1$, it follows that $L_\Delta(t)=4^t$. Let $L_v(t)$ and
$L_e(t)$ be the respective number of nodes and edges created at step
$t$. Notice that each triangle in $K(t-1)$ will lead to an addition
of six new nodes and nine new edges at step $t$, then one can easily
obtain the following relations: $L_v(t)=6\,L_\Delta(t-1)=6\times
4^{t-1}$ and $L_e(t)=9\,L_\Delta(t-1)=9\times 4^{t-1}$ for arbitrary
$t>0$. From these results, we can compute the order and the size of
Koch network. The total number of vertices $N_t$ and edges $E_t$
present at step $t$ is
\begin{equation}\label{Nt}
N_t=\sum_{t_i=0}^{t}L_v(t_i)=2\times 4^{t}+1
\end{equation}
and
\begin{equation}\label{Et}
E_t =\sum_{t_i=0}^{t}L_e(t_i)=3\times 4^{t},
\end{equation}
respectively. Thus, the average degree is
\begin{equation}\label{AveDeg}
\langle k \rangle =\frac{2E_t}{N_t} = \frac{6\times 4^{t}}{2\times
4^{t}+1},
\end{equation}
which is approximately $3$ for large $t$, showing that Koch network
is sparse as most real systems.

\section{\label{sec:topo}Structural properties of Koch network}

Now we study some relevant characteristics of Koch network $K(t)$,
focusing on degree distribution, clustering coefficient, average
path length, and degree correlations.

\subsection{Degree distribution}

We define $k_i(t)$ as the degree of a node $i$ at time $t$. When
node $i$ is added to the network at step $t_i$ ($t_i\geq 0$), it has
a degree of $2$, viz., $k_i(t_i)=2$. To determine $k_i(t)$, we first
determine the number of triangles involving node $i$ at step $t$
that is represented by $L_\Delta(i,t)$. These triangles will create
new nodes connected to the node $i$ at step $t+1$. Then at step
$t_i$, $L_\Delta(i, t_i)=1$. By construction,
$L_\Delta(i,t)=2\,L_\Delta(i,t-1)$. Since $L_\Delta(i, t_i)=1$, one
can derive $L_\Delta(i,t)=2^{t-t_{i}}$. Note that the relation
between $k_i(t)$ and $L_\Delta(i,t)$ satisfies:
\begin{equation}\label{ki}
k_i(t)=2\,L_\Delta(i,t)=2^{t+1-t_{i}}.
\end{equation}
In this way, at time $t$ the degree of node $i$ has been computed
explicitly. From Eq.~(\ref{ki}), one can see that at each step the
degree of a node doubles, i.e.,
\begin{equation}\label{ki2}
k_i(t)=2\,k_i(t-1).
\end{equation}

Equation~(\ref{ki}) shows that the degree spectrum of Koch network
is discrete. It follows that the cumulative degree
distribution~\cite{Ne03} is given by
\begin{equation}\label{pcumk}
P_{\rm cum}(k)=\frac{1}{N_t}\,\sum_{\tau \leq t_i}L_v(\tau)
={2\times 4^{t_i}+1 \over 2\times 4^{t}+1}.
\end{equation}
Substituting for $t_i$ in this expression using $t_i=t+1-\frac{\ln
k}{\ln 2}$ gives
\begin{equation}
P_{\rm cum}(k)=\frac{2\times 4^{t}\times 4 \times
k^{-(\ln4/\ln2)}+1}{2\times 4^{t}+1}.
\end{equation}
When $t$ is large enough, one can obtain
\begin{equation}\label{gammak}
P_{\rm cum}(k)=4 \times k^{-2}.
\end{equation}
So the degree distribution follows a power-law form with the
exponent $\gamma=3$. Note that this exponent of degree distribution
is the same as that of the Barab\'asi-Albert (BA)
model~\cite{BaAl99}.

Before closing this section, we compute another quantity $\langle
k^2 \rangle$, i.e., the fluctuations of the connectivity
distribution, which is useful for the calculation of Pearson
correlation coefficient that will be discussed in the following
text. The quantity $\langle k^2 \rangle$ is given by
\begin{eqnarray}\label{Ki21}
\langle k^{2}
\rangle=\frac{1}{N_t}\sum_{t_i=0}^{t}L_v(t_i)\left[k(t_i,t)\right]^{2},
\end{eqnarray}
where $k(t_i,t)$ is the degree of a node at step $t$ which was
generated at step $t_i$. Combining previously obtained results, we
find
\begin{eqnarray}\label{Ki22}
\langle k^{2} \rangle=\frac{2\times 4^t(3t+6)}{2\times 4^t+1}.
\end{eqnarray}

\subsection{Clustering coefficient}

The clustering coefficient~\cite{WaSt98} of a node $i$ with a degree
$k_i$ is given by $C_i =2e_i/[k_i(k_i-1)]$, where $e_i$ is the
number of existing triangles attached to node $i$, and
$k_i(k_i-1)/2$ is the total number of possible triangles including
$i$. Using the connection rules, it is straightforward to calculate
analytically the clustering coefficient $C(k)$ for a single node
with degree $k$. In the preceding section, we have obtained
$2\,e_{i}=k_{i}$ for all nodes at all steps. So there is a
one-to-one correspondence between the clustering coefficient of a
node and its degree. For a node of degree $k$, we have
\begin{equation}\label{Ck}
C(k)=\frac{1}{k-1},
\end{equation}
which is inversely proportional to $k$ in the limit of large $k$.
The scaling of $C(k)\sim k^{-1}$ has been observed in many
real-world scale-free networks~\cite{RaBa03}.

\begin{figure}
\begin{center}
\includegraphics[width=0.4\linewidth,trim=110 30 120 20]{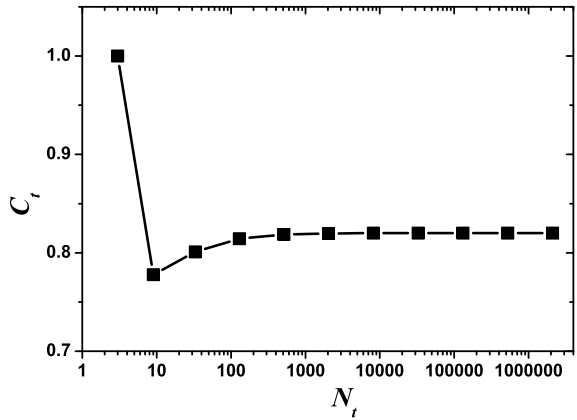} \
\end{center}
\caption[kurzform]{\label{clustering} Semilogarithmic plot of
average clustering coefficient $C_t$ versus network order $N_{t}$.}
\end{figure}

After $t$ generation evolutions, the clustering coefficient $C_t$ of
the whole network, defined as the average of $C_i\,^{'}s$ over all
nodes in the network, is given by
\begin{equation}\label{ACC}
C_t=
    \frac{1}{N_{t}}\sum_{r=0}^{t}
    \left [\frac{1}{M_r-1} L_v(r)\right],
\end{equation}
where the sum runs over all the nodes and $M_r$ is the degree of
those nodes created at step $r$, which is given by Eq.~(\ref{ki}).
In the limit of large $N_{t}$, Eq.~(\ref{ACC}) converges to a
nonzero value $C=0.82008$, as shown in Fig.~\ref{clustering}.
Therefore, the Koch network is highly clustered.

\subsection{Average path length}

\begin{figure}
\begin{center}
\includegraphics[width=0.40\linewidth,trim=110 30 120 20]{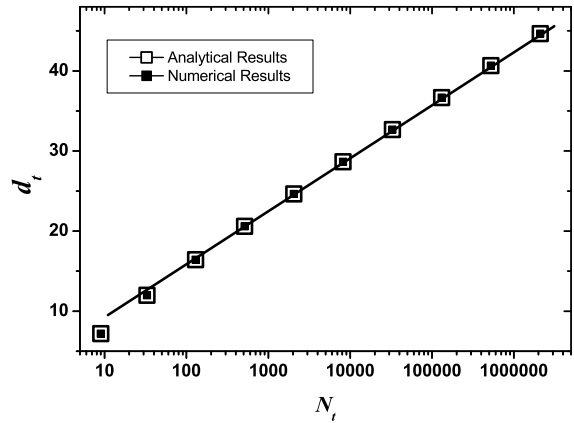}
\end{center}
\caption[kurzform]{\label{distance} Average path length $d_{t}$
versus network order $N_{t}$ on a semilogarithmic scale. The solid
line is a guide to the eye.}
\end{figure}

Let $d_t$ denote the APL of the Koch network $K(t)$. Since the Koch
network is self-similar, the APL can be computed analytically to
obtain an explicit formula, by using a method similar to but
different from those in Refs.~\cite{HiBe06,ZhChZhFaGuZo08}. We
represent all the shortest path lengths of $K(t)$ as a matrix in
which the entry $d_{ij}$ is the shortest distance from node $i$ to
node $j$, then $d_t$ is defined as the mean of $d_{ij}$ over all
couples of nodes,
\begin{equation}\label{eq:app1}
  d_{t} = \frac{D_t}{N_t(N_t-1)/2}\,,
\end{equation}
where
\begin{equation}\label{eq:app2}
  D_t = \sum_{\stackrel{i \in K(t),\, j \in K(t)}{i \neq j}} d_{ij}
\end{equation}
denotes the sum of the shortest path length between two nodes over
all pairs. It should be mentioned that in Eq.~(\ref{eq:app2}), for a
couple of nodes $i$ and $j$ ($i\neq j$), we only count $d_{ij}$ or
$d_{ji}$, not both. In the Appendix, we provide the detailed
derivation for the APL. The obtained analytical expression for $d_t$
is
\begin{equation}\label{APL}
d_t = \frac {4+14\times4^t+ 12t\times4^t}{3\,(4^t+1)},
\end{equation}
which approximates $4t$ in the infinite $t$, implying that the APL
shows a logarithmic scaling with the network order. Therefore, the
Koch network exhibits a small-world behavior. We have checked our
analytical result against numerical calculations for different
network orders up to $t=10$ which corresponds to $N_{10}=1\,048\,
577$. In all the cases we obtain a complete agreement between our
theoretical formula and the results of numerical investigation (see
Fig.~\ref{distance}).

\subsection{Degree correlations}

Degree correlations are a particularly interesting subject in the
field of network science~\cite{MsSn02}, because they can give rise
to some interesting network structure effects. An interesting
quantity related to degree correlations is the average degree of the
nearest neighbors for nodes with degree $k$, denoted as $k_{\rm
nn}(k)$, which is a function of node degree $k$~\cite{PaVaVe01}.
When $k_{\rm nn}(k)$ increases with $k$, it means that nodes have a
tendency to connect to nodes with a similar or larger degree. In
this case the network is defined as assortative~\cite{Newman02}. In
contrast, if $k_{\rm nn}(k)$ is decreasing with $k$, which implies
that nodes of large degree are likely to have near neighbors with
small degree, then the network is said to be disassortative. If
correlations are absent, $k_{\rm nn}(k)={\rm const}$.

We can exactly calculate $k_{\rm nn}(k)$ for Koch network using
Eqs.~(\ref{ki}) and (\ref{ki2}) to work out how many links are made
at a particular step to nodes with a particular degree. By
construction, we have the following
expression~\cite{DoMa05,ZhZhZoChGu07}:

\begin{eqnarray}\label{Knn1}
k_{\rm nn}(k)&=&{1\over L_v(t_i) k(t_i,t)}\Bigg(
  \sum_{t'_i=0}^{t'_i=t_i-1} L_v(t'_i) k(t'_i,t_i-1)k(t'_i,t) \nonumber\\
 &\quad& +\sum_{t'_i=t_i+1}^{t'_i=t} L_v(t_i) k(t_i,t'_i-1)
 k(t'_i,t)\Bigg)+1
\end{eqnarray}
for $k=2^{t+1-t_{i}}$. Here the first sum on the right-hand side
accounts for the links made to nodes with larger degree (i.e.,
$t'_i<t_i$) when the node was generated at $t_i$. The second sum
describes the links made to the current smallest degree nodes at
each step $t'_i>t_i$. The last term 1 accounts for the link
connected to the simultaneously emerging node. After some algebraic
manipulations, we obtain exactly
\begin{eqnarray}\label{knn3}
k_{\rm nn}(k)=t+2.
\end{eqnarray}
Therefore, two node correlations do not depend on the degree. On the
other hand, Eq.~(\ref{knn3}) shows that for large $t$, $k_{\rm
nn}(k)$ is approximately a logarithmic function of the network order
$N_t$, namely $k_{\rm nn}(k)\sim \ln N_t$. Note that the same
behavior has also been observed in the BA model~\cite{VapaVe02}.

Degree correlations can be also described by a Pearson correlation
coefficient $r$ of degrees at either end of a link. It is defined as
\cite{Newman02,DoMa05,RaDoPa04}
\begin{equation}\label{Pearson}
r={\langle k\rangle\langle k^2 k_{\rm nn}(k)\rangle -
    \langle k^2\rangle^2 \over
    \langle k\rangle \langle k^3\rangle - \langle k^2\rangle^2}.
\end{equation}
If the network is uncorrelated, the correlation coefficient equals
zero. Disassortative networks have $r<0$, while assortative graphs
have a value of $r>0$. Substituting Eqs. (\ref{AveDeg}),
(\ref{Ki22}), and (\ref{knn3}) into Eq. (\ref{Pearson}), we can
easily see that for arbitrary $t\geq 0$, the numerator of Eq.
(\ref{Pearson}) is always equal to zero. Thereby, $r$ also equals
zero, which again indicates that Koch network shows the absence of
degree correlations.

\section{\label{sec:standwalk}Random walks}

As addressed in Sec.~\ref{sec:topo}, the Koch network exhibits
exclusive topological properties not simultaneously shared by other
networks. Thus, it is worthwhile to study dynamical processes
occurring on the network. In this section we consider simple random
walks on the Koch network defined by a walker such that at each step
the walker, located on a given node, moves to any of its nearest
neighbors with equal probabilities.

\subsection{Scaling efficiency}

We follow the concept of {\em scaling efficiency} introduced
in~\cite{Bobe05}. Denote by $T_{ij}$ the first-passage time (FPT)
between two nodes $i$ and $j$ in a network. Let $T_{ii}$ be the mean
time for a walker returning to a node $i$ for the first time after
the walker has left it. When the network order grows from $N$ to
$gN$, one expects that in the infinite limit of $N$
\begin{equation}
T_{ij}(gN)\sim g^{\delta_{ij}}T_{ij}(N),
\end{equation}
where $\delta_{ij}$ is defined as the scaling efficiency exponent.
An analogous relation for $T_{ii}$ defines an exponent
$\delta_{ii}$.

One can confine the scaling efficiency in the nodes already existing
in the network before growth. Let $T'_{ij}(gN)$ be the mean
first-passage time in the network under consideration, averaged over
the \emph{original} class of nodes (before growth). Then the
\emph{restricted} scaling efficiency exponent $\lambda_{ij}$ is
defined by the relation
\begin{equation}
T'_{ij}(gN)\sim g^{\lambda_{ij}}T_{ij}(N).
\end{equation}
Similarly, we can define $\lambda_{ii}$.

After introducing the concepts, in the following we will investigate
random walks on the Koch network following a similar but obviously
different method used in~\cite{Bobe05,ZhZhZhYiGu09}.

\subsection{First-passage time for old nodes}

\begin{figure}
\begin{center}
\includegraphics[width=.65\linewidth,trim=60 30 60 10]{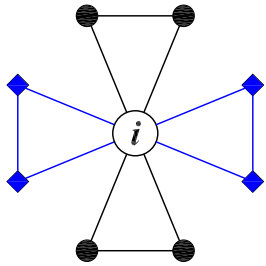}
\end{center}
\caption{(Color online) Growth of trapping time in going from
$K_{t}$ to $K_{t+1}$. Node $i \in K_{t}$ has $k_i$ neighbor nodes in
generation $t$ ($\bullet$) and $k_i$ new neighbor nodes in
generation $t+1$ ($\blacksquare$). A new neighbor of node $i$ has a
degree of 2, and is simultaneously linked to another new neighbor of
$i$.} \label{Trap}
\end{figure}

Consider an arbitrary node $i$ in the Koch network $K(t)$ after $t$
generation evolution. Note that for the sake of simplicity, we also
denote $K(t)$ by $K_t$, and both denotations will be used
alternatively in the following text. From Eq.~(\ref{ki2}), we know
that upon growth of the network to generation $t+1$, the degree
$k_i$ of node $i$ doubles, that is to say, it increases from $k_i$
to $2k_i$. Let the FPT for going from node $i$ to any of the $k_i$
old neighbors be $T$ and let the FPT for going from any of the $k_i$
new neighbors to one of the $k_i$ old neighbors be $A$. Then we can
establish the following equations (see Fig.~\ref{Trap}):
\begin{eqnarray}\label{FPT1}
\left\{
\begin{array}{ccc}
T&=&\frac{1}{2}+\frac{1}{2}(1+A),\\
A&=&\frac{1}{2}(1+T)+\frac{1}{2}(1+A),
 \end{array}
 \right.
\end{eqnarray}
which leads to $T=4$. Therefore, the passage time from any node $i$
($i \in K_t$) to any node $j$ ($j\in K_{t+1}$) increases four times,
on average, upon growth of the network to generation $t+1$, i.e.,
\begin{equation}\label{FPT2}
 T'_{ij}(N_{t+1})=4\,T_{ij}(N_t).
\end{equation}
For explanation, see Refs.~\cite{Bobe05,HaBe87} and related
references therein. Since the network order approximately grows by
four times in the large $t$ limit [see Eq.~(\ref{Nt})]. This
indicates that the scaling efficiency exponent for old nodes is
$\lambda_{ij}=1$, which is the same as that of the recursive
scale-free tree addressed in Ref.~\cite{Bobe05}.

Next we continue to consider the return of FPT to node $i$. Denote
by $T'_{ii}$ the FPT for returning to node $i$ in $K_{t+1}$. Denote
by $T'_{ji}$ the FPT from $j$---an old neighbor of $i$ ($i \in
K_{t}$)---to $i$, in $K_{t+1}$. Analogously, denote by $T_{ii}$ the
FPT for returning to $i$ in $K_{t}$ and denote by $T_{ji}$ the FPT
from the same neighbor $j$, to $i$, in $K_{t}$. For $K_{t}$, we have
\begin{equation}\label{FPT3}
 T_{ii}=\frac{1}{k_i}\sum_{j \in \Omega_i(t)}(1+T_{ji}),
\end{equation}
where $\Omega_i(t)$ is the set of neighbors of node $i$, which
belongs to $K_{t}$. On the other hand, for $K_{t+1}$,
\begin{equation}\label{FPT4}
T'_{ii}=\frac{1}{2}\times
3+\left(1-\frac{1}{2}\right)\frac{1}{k_i}\sum_{j \in
\Omega_i(t)}(1+T'_{ji}).
\end{equation}
The first term on the rhs of Eq.~(\ref{FPT4}) accounts for the
process in which the walker moves from node $i$ to its new neighbors
and back. Since among all neighbors of node $i$, half of them are
new, which is obvious from Eq.~(\ref{ki2}), such a process occurs
with a probability of $\frac{1}{2}$ and takes three time steps. The
second term on the rhs interprets the process where the walker steps
from $i$ to one of the old neighbors $j$ previously existing in
$K_{t}$ and back; this process happens with the complimentary
probability $(1-\frac{1}{2})$.

Using Eq.~(\ref{FPT2}) to simplify Eq.~(\ref{FPT4}), we can obtain
\begin{equation}\label{FPT5}
T'_{ii}=2\,T_{ii}=4^{1/2}\,T_{ii}.
\end{equation}
In other words,
\begin{equation}\label{FPT6}
T'_{ii}(N_{t+1})=4^{1/2}\,T_{ii}(N_{t}).
\end{equation}
Thus, the scaling efficiency exponent $\lambda_{ii}=\frac{1}{2}$,
which is less than 1. Recall that for the recursive scale-free tree,
its scaling efficiency exponent is $\lambda_{ii}=1-\frac{\ln 2}{\ln
3}<\frac{1}{2}$~\cite{Bobe05}, which means that in the Koch network
it is more difficult for the walker to return to the origin than in
the recursive scale-free tree, when the networks grow in size.

\subsection{First-passage time for all nodes}

We continue to compute $T_{j'j'}$, which is the FPT to return to a
new node $j'\in K_{t}$ that is a neighbor of node $i \in K_{t-1}$.
Notice that when $j'$ was generated, another node $j''$ emerged
simultaneously, connected to $i$ and $j'$ (see Fig.~\ref{Trap}).
Denote by $T_1$ the FPT from $i$ to $j'$ and denote by $B$ the FPT
to return to $i$ (starting off from $i$) without ever visiting $j'$
and $j''$. Then we have
\begin{equation}\label{FPT7}
T_{j'j'}=\frac{1}{2}(1+T_1)+\frac{1}{2}(1+T_{j''j'}),
\end{equation}
\begin{equation}\label{FPT71}
T_{j''j'}=\frac{1}{2}\times 1+\frac{1}{2}(1+T_1),
\end{equation}
and
\begin{equation}\label{FPT8}
T_1=\frac{1}{k_i}+\frac{1}{k_i}(1+T_{j''j'})+\frac{k_i-2}{k_i}(B+T_1).
\end{equation}
Equation~(\ref{FPT8}) can be interpreted as follows: with
probability $\frac{1}{k_i}$ ($k_i$ being the degree of node $i$ in
$K_{t}$), the walker starting from node $i$ would take one time step
to go to node $j'$; with probability $\frac{1}{k_i}$, the walker
takes one time step to move to node $j''$ then takes time
$T_{j''j'}$ to reach $j'$; and with the remaining probability
$\frac{k_i-2}{k_i}$, the walker chooses uniformly a neighbor node
except $j'$ and $j''$ and spends on average time $B$ in returning to
$i$ then takes time $T_1$ to arrive at node $j'$.

In order to close Eqs.~(\ref{FPT7}) and~(\ref{FPT8}), we express the
FPT to return to $i$ as
\begin{equation}\label{FPT9}
T_{ii}(N_{t})=\frac{1}{k_i}\times 3+\frac{1}{k_i}\times
3+\frac{k_i-2}{k_i}B.
\end{equation}
Eliminating $T_1$, $T_{j''j'}$, and $B$, we obtain
\begin{equation}\label{FPT10}
T_{j'j'}(N_{t})=\frac{k_i}{2}\,T_{ii}(N_t).
\end{equation}
Combining Eqs.~(\ref{ki}), (\ref{FPT5}), and~(\ref{FPT10}), we have
\begin{equation}\label{FPT11}
T_{j'j'}(N_{t})=3\times 4^{t}\sim 3\,N_t.
\end{equation}
Thus, in spite of the fact that simultaneously emerging new nodes
are linked to different nodes with various degrees, they have the
same mean return time. Iterating Eqs.~(\ref{FPT5})
and~(\ref{FPT10}), we have that in $K_{t}$ there are $L_v(\epsilon)$
($0\leq \epsilon \leq t$) nodes with $T_{ii}=2^{t+1+\epsilon}$.
This, together with Eqs. (\ref{Et}) and (\ref{ki}), means that for
an arbitrary node $i$ (born at step $t_i$) with degree
$k_i(t)=2^{t+1-t_i}$ at time $t$, the FPT to return to $i$ is
$T_{ii}=\frac{2\,E_t}{k_i(t)}$. Note that a similar expression has
been obtained for the periodic lattices~\cite{Hu95} and random
networks~\cite{NoRi04} by using a different derivation method. Thus,
the mean return time $T_{ii}$ is not reliant on the details of the
global structure of Koch network. It only relies on the network size
and the connectivity of the node: the larger the degree of the node,
the smaller the FPT to return.

Taking the average of $T_{ii}$ over all nodes in $K_{t}$ leads to
\begin{eqnarray}\label{FPT12}
\langle T_{ii}\rangle_t&=&\frac{1}{2\times 4^{t}+1}\big (3\times
2^{t+1}+6\times 2^{t+2}+6\times 4\times2^{t+2} \nonumber \\&\quad&
\quad\quad\quad\quad\quad\quad\quad+6\times
4^{t-1}\times2^{t+1+t}\big )\nonumber \\
&=& \frac{1}{2\times 4^{t}+1}\left(\frac{24}{7}\times 16^t
+\frac{18}{7}\times 2^t \right).
\end{eqnarray}
For infinite $t$, $\langle T_{ii}\rangle_t \approx
\frac{12}{7}\times4^{t} \sim N_t $, implying that $\delta_{ii}=1$,
an efficiency scaling identical to that of the recursive scale-free
tree~\cite{Bobe05}. Therefore, the scalings of new nodes play a
dominant role in the average of mean return time for all nodes.

Next we calculate $T_{ij}$ in $K_{t}$, which is FPT from an
arbitrary node $i$ to another node $j$. Since each newly created
node has a degree of 2 and is linked to an old node and a
simultaneously emerging new node, and these three nodes form a
triangle, the FPT $T_{i'j}$ from node $i'$---a new neighbor of the
old node $i$---to $j$ equals $T_{ij}$ plus 2 (i.e., $T_{i'i}$) and
thus has little effect on the scaling when the network order $N$ is
very large. Therefore, we need only to consider FPT $T_{ij'}$ from
$i$ to $j'$---a new neighbor of $j$, which can be expressed as
\begin{equation}\label{FPT13}
T_{ij'}=T_{ij}+T_{jj'}.
\end{equation}
Suppose that when $j'$ was born, it connected to node $j$ and a
simultaneously emerging node $j''$ (by construction, $j''$ was also
linked to node $j$), then we have
\begin{equation}\label{FPT14}
T_{j'j'}=\frac{1}{2}(1+T_{jj'})+\frac{1}{2}(1+T_{j''j'}),
\end{equation}
and
\begin{equation}\label{FPT14a}
T_{j''j'}=\frac{1}{2}+\frac{1}{2}(1+T_{jj'}).
\end{equation}
Inserting Eq.~(\ref{FPT14a}) to Eq.~(\ref{FPT14}), we obtain
\begin{equation}\label{FPT14c}
T_{jj'}=\frac{4}{3}\,T_{j'j'}-2.
\end{equation}
Substituting Eq.~(\ref{FPT14c}) and Eq.~(\ref{FPT11}) for $T_{j'j'}$
into Eq.~(\ref{FPT13}) results in
\begin{equation}\label{FPT15}
T_{ij'}=T_{ij}+\frac{4}{3}\,T_{j'j'}-2\sim N_t,
\end{equation}
where Eq.~(\ref{FPT2}) has been used. Therefore, we have
\begin{equation}\label{FPT16}
\langle T_{ij}\rangle_t\sim N_t,
\end{equation}
which shows that the mean transit time between arbitrary pairs of
nodes is proportional to the network order. Equation~(\ref{FPT16})
also reveals that $\delta_{ij}$ is a constant 1. Notice that the
linear scaling of the average traverse time with network order has
been previously obtained by numerical simulations for the Apollonian
networks~\cite{HuXuWuWa06} and the pseudofractal scale-free
web~\cite{Bobe05}, both of which have been well
studied~\cite{AnHeAnSi05,DoMa05,ZhCoFeRo06,ZhRoZh06,ZhGuXiQiZh09,KaHiBe09,DoGoMe02,CoFeRa04,ZhZhCh07,ZhRoZh07,RoHaAv07,ZhQiZhXiGu09}.

\section{Random walk with a trap}

In the preceding section, we have obtained the scaling between the
MFPT and the network order. Although scaling theory is central in
studying diffusive process on a variety of deterministic media, it
has been noted that scaling laws do not provide a complete picture
of dynamic phenomena on deterministic media, e.g., regular
fractals~\cite{Gi96}, and exact relationships are useful. In this
section we study the trapping problem of a simple unbiased Markovian
random walk of a particle on network $K_t$ in the presence of a trap
or a perfect absorber located on a given node, which absorbs all
particles visiting it. Our aim is to give some further insight into
the trapping process and the independence of the average trapping
time (TT) on the size of the underlying system, by providing a
rigorous solution of the mean trapping time (MTT) as a function of
the network order.

\subsection{Formulating the trapping problem}

For the convenience of description, we distinguish different nodes
by labeling all the nodes belonging to $K_t$ in the following way.
The initial three nodes in $K_0$ are labeled as 1, 2, and 3,
respectively. In each new generation, only the new nodes created at
this generation are labeled, while the labels of all old nodes
remain unchanged, i.e., we label new nodes as $M+1$, $M+2$,
$\ldots$, $M+\Delta M$, where $M$ is the total number of the
pre-existing nodes and $\Delta M$ is the number of newly created
nodes. Eventually, every node is labeled by a unique integer, at
time $t$ all nodes are labeled from 1 to $N_t=2\times 4^{t}+1$ (see
Fig.~\ref{network}).


Before proceeding further, we give the following definitions. Let
$A_{i j}$ be an element of the adjacency matrix $\textbf{A}_{t}$ of
network $K_t$ such that $A_{i j}=1$ if nodes $i$ and $j$ are
connected by an edge and $A_{i j}=0$ otherwise. Thus, the degree of
a vertex $i$ in $K_t$ is $d_{v_i}=\sum_{j} A_{i j}$. And the
diagonal degree matrix $\textbf{D}_{t}$ of  $K_t$ is defined by
$\textbf{D}_{t}=\text{diag} (d_{ v_1}, d_{ v_2},\ldots, d_{ v_i},
\ldots, d_{v_{N_t}})$. Finally, we define
$\textbf{W}_{t}=\textbf{A}_{t}\cdot\textbf{D}_{t}^{-1}$, where
$\textbf{D}_{t}^{-1}$ is the inverse matrix of $\textbf{D}_{t}$,
then the normalized Laplacian matrix of network $K_t$ is
$\textbf{L}_{t}=\textbf{I}_{t}-\textbf{W}_{t}$, in which
$\textbf{I}_{t}$ is an identity matrix with order $N_t \times N_t$.

We locate the trap at node 1 (due to the symmetry, the trap can be
also located at node 2 or 3, which does not have any effect on
MFPT), denoted as $i_T$. Note that the particular selection we made
for the trap location makes the analytical computation process (that
will be shown in detail in the following text) easily iterated as we
can identify the trap node $i_T$ since the first generation. At each
time step (taken to be unity), the walker selects uniformly among
the neighbors of the current node (excluding the trap) and takes a
step to one of them. Since node 1 is one of the three nodes with the
largest degree, it is easily seen that in the presence of the trap
$i_T$ fixed on node 1, the walker will be inevitably
absorbed~\cite{Bobe05}.

This trapping process can be described by specifying the set of
transition probabilities $W_{ij}$ for the particle of going from
node $i$ (except the trap $i_T$) to node $j$. We can regard $W_{ij}$
as an element of the matrix $\textbf{W}_{t}^{T}$, which is a
submatrix of $\textbf{W}_{t}$ with the row and the column
corresponding to trap being removed. That is to say,
$\textbf{W}_{t}^{T}$ is a $(N_t-1)$-order submatrix of
$\textbf{W}_{t}$ with the first row and column being deleted.
Similarly, one can define $\textbf{A}_{t}^{T}$,
$\textbf{D}_{t}^{T}$, $\textbf{I}_{t}^{T}$ and $\textbf{L}_{t}^{T}$,
thus we have
$\textbf{W}_{t}^{T}=\textbf{A}_{t}^{T}\cdot(\textbf{D}_{t}^{T})^{-1}$
and $\textbf{L}_{t}^{T}=\textbf{I}_{t}^{T}-\textbf{W}_{t}^{T}$. The
inverse of $\textbf{L}_{t}^{T}$, $(\textbf{L}_{t}^{T})^{-1}$, is the
fundamental matrix of the Markovian chain representing the unbiased
random walk with a trap.

An interesting quantity related to the trapping process is the mean
residence time (MRT), which is the mean time that a random walker
spends at a given site prior to being trapped. In fact, the MRT is
the mean number of visitations of a given site by the walker before
trapping occurs, and finding the MRT at node $i$, starting from node
$j$, is equivalent to finding the element $L_{i,j}^{-1}$ of the
fundamental matrix~\cite{KeSn76}.

Another quantity of interest is the TT. In network $K_t$, the
trapping time $T_i^t$ of a given site $i$ is the expected time for a
walker starting from $i$ to first reach the trap. By definition,
trapping time $T_i^t$ is the sum of the MRTs over all nodes except
$i_T$, i.e.,
\begin{equation}\label{MFPT4}
T_i^t=\sum_{j=2}^{N_t}{L_{ij}^{-1}}.
\end{equation}

Then, the MTT or the mean first-passage time (MFPT), $\langle T
\rangle_t$, which is the average of $T_i^t$ over all initial nodes
distributed uniformly over nodes in $K_t$ other than the trap, is
given by
\begin{equation}\label{MFPT5}
 \langle T
\rangle_t=\frac{1}{N_t-1}\sum_{i=2}^{N_t}
T_i^t=\frac{1}{N_t-1}\sum_{i=2}^{N_t}\sum_{j=2}^{N_t}{L_{ij}^{-1}}.
\end{equation}

The quantities of TT and MTT are very important since they measure
the efficiency of the trapping process: the smaller the two
quantities, the higher the efficiency, and vice versa.
Equations~(\ref{MFPT4}) and~(\ref{MFPT5}) show that the problem of
calculating $T_i^t$ and $\langle T \rangle_t$ is reduced to finding
the sum of elements of matrix $(\textbf{L}_{t}^{T})^{-1}$. In
Tables~\ref{tab:AMTA1} and~\ref{tab:AMTA2}, we list separately
$T_i^t$ of some nodes and $\langle T \rangle_t$ for different
network orders up to $t=6$. From Table~\ref{tab:AMTA1}, one can
easily observe that for a given node $i$, the relation
$T_i^{t+1}=4\,T_i^t$ holds. That is to say, upon growth of Koch
network from $t$ generation to generation $t+1$, the trapping time
to first reach the trap increases by a factor of 4, which is
consistent with Eq.~(\ref{FPT2}). This scaling relation is a basic
character of the trapping process on the Koch network, which will be
useful for deriving the formula of MTT that will be given in the
following section.

\begin{table*}
\caption{Trapping time $T_i^t$ for a random walker starting from
node $i$ on $K_t$ for various $t$. Notice that owing to the obvious
symmetry, nodes in a parentheses are equivalent, since they have the
same trapping time. All the values are calculated straightforwardly
from Eq.~(\ref{MFPT4}).} \label{tab:AMTA1}
\begin{center}
\begin{tabular}{l|cccccccccc}
\hline \hline  $t\backslash i$ & \quad$(2,3)$\quad \quad&\quad $(4,5)$\quad\quad & $(6,7,8,9)$ & $(10,11,12,13)$ & $(14,15,16,17,18,19,20)$ & $(22,23,24,25)$ & $(26,27,28,29,30,31,32,33)$    \\
            \hline
            0 & $2$ \\
            1 & $8$ & $2$ & $10$      \\
            2 & $32$ & $8$ & $40$ & $2$ & $34$ & $10$ & $42$  \\
            3 & $128$  & $32$ & $160$ & $8$ & $136$ & $40$ & $168$ \\
            4 & $512$ & $128$ & $640$ & $32$ & $544$ & $160$ & $672$  \\
            5 & $2048$ & $512$ & $2560$ & $128$ & $2176$ & $640$ & $2688$ \\
            6 & $8192$ & $2048$ & $10240$ & $512$ & $8704$ & $2560$ & $10752$ \\
\hline \hline
\end{tabular}
\end{center}
\end{table*}

Notice that the order of matrix $(\textbf{L}_{t}^{T})^{-1}$ is
$(N_t-1)\times (N_t-1)$, where $N_t$ increases exponentially with
$t$, as shown in Eq.~(\ref{Nt}). Thus, for large $t$, the
computation of the TT and the MTT from Eqs.~(\ref{MFPT4})
and~(\ref{MFPT5}) is prohibitively time and memory consuming, making
it difficult to obtain $T_i^t$ and $\langle T \rangle_t$ through
direct calculation for large network; one can compute directly the
MFPT only for the first several generations. However, the recursive
construction of Koch network allows one to compute analytically the
MTT to achieve a closed-form solution; the derivation details of
which will be given in next section.

\begin{table}
\caption{Mean trapping time obtained by direct calculation from
Eq.~(\ref{MFPT5}).} \label{tab:AMTA2}
\begin{center}
\begin{tabular}{ccccc}
\hline \hline
                \quad $t$\quad\quad & \quad $N_t$ \quad\quad  & \quad\quad$\sum_{i=2}^{N_t} T_i^t$\quad\quad &\quad \quad$\langle T
\rangle_t$ \quad\quad  \\
                \hline
                0   &  $3$      &   $4$ &     $4/2$ \\
                1   &  $9$      &   $60$ &    $60/8$  \\
                2   &  $33$     &   $896$   &   $896/32$    \\
                3   &  $129$     &   $13888$    &   $13888/128$ \\
                4   &  $513$    &   $219648$    &   $219648/512$    \\
                5   &  $2049$   &   $3501056$  &   $3501056/2048$ \\
                6   &  $8193$   &   $55951360$  &   $55951360/8192$ \\
\hline \hline
\end{tabular}
\end{center}
\end{table}

\subsection{Analytical solution for mean trapping time}

We now determine the average of the mean time to absorption, aiming
to derive an exact solution. We represent the set of nodes in
$K_{t}$ as $\Lambda_t$ and denote the set of nodes created at
generation $t$ by $\overline{\Lambda}_t$. Thus we have
$\Lambda_t=\overline{\Lambda}_t\cup\Lambda_{t-1}$. For the
convenience of computation, we define the following quantities for
$m \leq t$:
\begin{equation}\label{MFPT7}
 T_{m,\text{sum}}^t=\sum_{i\in \Lambda_m} T_i^t
\end{equation}
and
\begin{equation}\label{MFPT8}
 \overline{T}_{m,\text{sum}}^t=\sum_{i\in \overline{\Lambda}_m}
 T_i^t.
\end{equation}
Then, we have
\begin{equation}\label{MFPT9}
 T_{t,\text{sum}}^t=T_{t-1,\text{sum}}^t+\overline{T}_{t,\text{sum}}^t.
\end{equation}
Next we will explicitly determine the quantity $T_{t,\text{sum}}^t$.
To this end, we should first determine
$\overline{T}_{t,\text{sum}}^t$.

We examine the mean time to absorption for the first several
generations of the Koch network. Obviously, for all $t \geq 0$,
$T_1^t=0$; for $t = 0$, it is a trivial case. We have
$T_2^0=T_3^0=2$. In the case of $t=1$, by construction of the Koch
network, it follows that
$T^{1}_4=\frac{1}{2}(1+T^{1}_1)+\frac{1}{2}(1+T^{1}_5)$,
$T^{1}_5=\frac{1}{2}(1+T^{1}_1)+\frac{1}{2}(1+T^{1}_4)$,
$T^{1}_6=\frac{1}{2}(1+T^{1}_2)+\frac{1}{2}(1+T^{1}_7)$,
$T^{1}_7=\frac{1}{2}(1+T^{1}_2)+\frac{1}{2}(1+T^{1}_6)$,
$T^{1}_8=\frac{1}{2}(1+T^{1}_3)+\frac{1}{2}(1+T^{1}_9)$, and
$T^{1}_9=\frac{1}{2}(1+T^{1}_3)+\frac{1}{2}(1+T^{1}_8)$. Thus,
\begin{eqnarray}\label{MFPT10}
\overline{T}_{1,\text{sum}}^1&=&\sum_{i\in
\overline{\Lambda}_1}T^{1}_i=T^{1}_4+T^{1}_5+T^{1}_6+T^{1}_7+T^{1}_8+T^{1}_9\nonumber\\
&=&12+2(T^{1}_1+ T^{1}_2+
T^{1}_3)=12+2\,\overline{T}_{0,\text{sum}}^1\,.
\end{eqnarray}
Similarly, for $t=2$ case, we have
\begin{eqnarray}\label{MFPT11}
\overline{T}_{2,\text{sum}}^2&=&\sum_{i\in
\overline{\Lambda}_2}T^{2}_i=\sum_{i=10}^{33}T^{2}_i\nonumber\\
&=&48+2^2\,(T^{2}_1+ T^{2}_2+
T^{2}_3)\nonumber\\&\quad&+2\,(T^{2}_4+ T^{2}_5+ T^{2}_6+T^{2}_7+
T^{2}_8+
T^{2}_9)\nonumber\\
&=&12\times
4^{1}+2^2\,\overline{T}_{0,\text{sum}}^2+2\,\overline{T}_{1,\text{sum}}^2\,.
\end{eqnarray}
Proceeding analogously, it is not difficult to derive that
\begin{eqnarray}\label{MFPT12}
\overline{T}_{t,\text{sum}}^t=12\times 4^{t-1}&+&2\,\overline{T}^{t}_{t-1,\text{sum}}+2^2\,\overline{T}^{t}_{t-2,\text{sum}}+\cdots \nonumber\\
&+&2^{t-1}\,\overline{T}^{t}_{1,\text{sum}}+2^{t}\,\overline{T}^{t}_{0,\text{sum}},
\end{eqnarray}
and
\begin{eqnarray}\label{MFPT13}
\overline{T}_{t+1,\text{sum}}^{t+1}=12\times 4^{t}&+&2\,\overline{T}^{t+1}_{t,\text{sum}}+2^2\,\overline{T}^{t+1}_{t-1,\text{sum}}+\cdots \nonumber\\
&+&2^{t}\,\overline{T}^{t+1}_{1,\text{sum}}+2^{t+1}\,\overline{T}^{n+1}_{0,\text{sum}}\,,
\end{eqnarray}
where $12\times 4^{t-1}$ and $12\times 4^{t}$ are indeed the double
of node numbers, which were generated at generations $t$ and $t+1$,
respectively. Equation~(\ref{MFPT13}) minus Eq.~(\ref{MFPT12}) times
8 and making use of the relation $T_i^{t+1}=4\,T_i^t$, one gets
\begin{equation}\label{MFPT14}
\overline{T}^{t+1}_{t+1,\text{sum}}-12\times
4^{t}=2\,\overline{T}^{t+1}_{t,\text{sum}}+8\,\big(\overline{T}^{t}_{t,\text{sum}}-12\times
4^{t-1}\big),
\end{equation}
which may be rewritten as
\begin{equation}\label{MFPT15}
\overline{T}^{t+1}_{t+1,\text{sum}}=16\,\overline{T}^{t}_{t,\text{sum}}-12\times
4^{t}.
\end{equation}
Using $\overline{T}_{1,\text{sum}}^1=44$, Eq.~(\ref{MFPT15}) is
solved inductively as
\begin{equation}\label{MFPT16}
\overline{T}^{t}_{t,\text{sum}}=5\times 2^{4t-1}+4^{t}\,.
\end{equation}
Substituting Eq.~(\ref{MFPT16}) for
$\overline{T}^{t}_{t,\text{sum}}$ into Eq.~(\ref{MFPT9}), we have
\begin{eqnarray}\label{MFPT17}
T_{t,\text{sum}}^t&=&T_{t-1,\text{sum}}^t+5\times 2^{4t-1}+4^{t}\nonumber\\
&=&4\,T_{t-1,\text{sum}}^{t-1}+5\times 2^{4t-1}+4^{t}\,.
\end{eqnarray}
Considering the initial condition $T_{0,\text{sum}}^0=4$,
Eq.~(\ref{MFPT17}) is resolved by induction to yield
\begin{equation}\label{MFPT18}
T_{t,\text{sum}}^t=\frac{4^t}{3}(10 \times 4^t + 3\,t+2) \,.
\end{equation}
Plugging the last expression into Eq.~(\ref{MFPT5}), we arrive at
the accurate formula for the average of the mean time to absorption
at the trap located at node 1 on the $t$th of the Koch network:
\begin{eqnarray}\label{MFPT19}
 \langle T
\rangle_t &=&\frac{1}{N_t-1}\sum_{i=2}^{N_t}
T_i=\frac{1}{N_t-1}T_{t,\text{sum}}^t \nonumber\\
&=& \frac{5}{3}\times4^t+\frac{t}{2}+\frac{1}{3}\,.
\end{eqnarray}

We continue to show how to represent mean trapping time as a
function of network order, with the aim of obtaining the scaling
between these two quantities. Recalling Eq.~(\ref{Nt}), we have
$4^{t}=\frac{N_t-1}{2}$ and $t=\log_4(N_t-1)-\frac{1}{2}$. These two
relations enable us to write Eq.~(\ref{MFPT19}) as
\begin{equation}\label{MFPT20}
 \langle T
\rangle_t
=\frac{5}{6}(N_t-1)+\frac{1}{2}\log_4(N_t-1)+\frac{1}{12}\,,
\end{equation}
from which it is easy to see that for large network (i.e.,
$N_t\rightarrow \infty$), the following expression holds:
\begin{equation}\label{MFPT21}
\langle T \rangle_t \approx \frac{5}{6}\,N_t\, .
\end{equation}
Thus, the mean trapping time grows linearly with increasing order of
the network, which is consistent with the conclusion obtained in the
preceding section.

\begin{figure}
\begin{center}
\includegraphics[width=0.40\linewidth,trim=115 40 120 40]{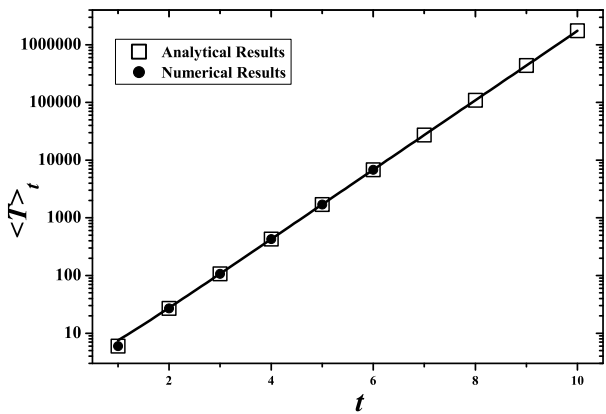}
\end{center}
\caption[kurzform]{\label{trap} Mean first-passage time $\langle T
\rangle_t$ versus $t$ on a semilogarithmic scale. The solid line
corresponds to the relation between $\langle T \rangle_t$ and
network order $N_{t}$: $\langle T \rangle_t \approx
\frac{5}{6}\,N_t=\frac{5}{6}\,(2\times 4^t+1)$ as given by
Eq.~(\ref{MFPT21}).}
\end{figure}

We have checked our analytical formulas, i.e.,
Eqs.~(\ref{MFPT19})-(\ref{MFPT21}), against numerical values quoted
in Table~\ref{tab:AMTA2}. For the range of $1 \leq t \leq 6$, the
values obtained from Eq.~(\ref{MFPT19}) or  Eq.~(\ref{MFPT20})
completely agree with those numerical results on the basis of the
direct calculation through Eq.~(\ref{MFPT5}) (see also
Fig.~\ref{trap} for comparison). This agreement serves as an
independent test of our theoretical formulae.

Notice that this linear scaling between the average trapping time
and the network order has been previously obtained for
three-dimensional regular lattice by using a method of generating
function~\cite{Mo69}. It is also interesting to stress that this
linear scaling is in contrast to the sub-linear scaling of mean
trapping time obtained for the two-dimensional Apollonian
network~\cite{ZhGuXiQiZh09} and the pseudofractal scale-free
web~\cite{ZhQiZhXiGu09}, in spite of the fact that they have a
similar topological
structure~\cite{AnHeAnSi05,ZhCoFeRo06,ZhRoZh06,DoGoMe02,CoFeRa04,ZhZhCh07}
as that of the Koch network. The reason for this disparity is worth
studying in the future.

\section{Conclusion}

In this paper, on the basis of the well-known Koch fractal we have
proposed a scale-free network that is called Koch network. We have
provided a detailed exact analysis of the topological features. We
have shown that the Koch network displays a rich structural
behavior: it is simultaneously scale free and small world, and has a
high clustering coefficient. In particular, we have shown that the
Koch network is an absolutely two-point uncorrelated network. The
especial structural characteristics make Koch network unique within
the class of scale-free networks.

The uniqueness of the topological features for Koch network
presented here makes it potentially interesting to study dynamical
processes occurring on the network. We have investigated the random
walk process and a particular trapping problem (with a trap located
at a hub node) on the Koch network. We have obtained analytically
the scaling efficiency exponents which are interesting to random
walks. We have shown analytically that the mean first-passage time
behaves linearly with the number of network nodes [see
Eqs.~(\ref{FPT16}) and~(\ref{MFPT21})], which indicates that despite
the presence of loops in the Koch network, the linear scaling of the
MFPT for random walks is similar to that of a scale-free
tree~\cite{Bobe05}. Our analytical result confirms a previous
conclusion obtained by numerical simulations that for scale-free
networks (with loops or not): the MFPT of random walks increases
linearly with the network order~\cite{HuXuWuWa06,Bobe05}.

\subsection*{Acknowledgment}

We thank Yichao Zhang and Ming Yin for their help. This research was
supported by the National Basic Research Program of China under
Grant No. 2007CB310806; the National Natural Science Foundation of
China under Grants No. 60704044, No. 60873040, and No. 60873070; the
Shanghai Leading Academic Discipline Project No. B114, and the
Program for New Century Excellent Talents in University of China
(Grant No. NCET-06-0376). W L Xie also acknowledges the support
provided by Hui-Chun Chin and Tsung-Dao Lee Chinese Undergraduate
Research Endowment (CURE).

\appendix

\section{Derivation of the average path length}

\begin{figure}
\begin{center}
\includegraphics[width=.7\linewidth,trim=100 0 100 0]{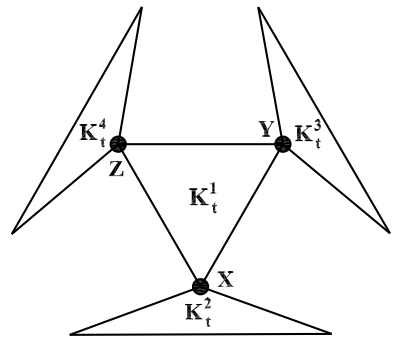}
\caption{Second construction method of Koch network that highlights
self-similarity. The graph after $t+1$ construction steps, $K(t+1)$,
is composed of four copies of $K(t)$ denoted as
    $K_t^{\theta}$ $(\theta=1,2,3,4)$, which are connected to one another as above.} \label{copy}
\end{center}
\end{figure}

The Koch network has a self-similar structure that allows one to
calculate $d_{t}$ analytically. The self-similar structure is
obvious from an equivalent network construction method: to obtain
$K(t+1)$, one can make four copies of $K(t)$ and join them at the
hub nodes. As shown in Fig.~\ref{copy}, network $K(t+1)$ may be
obtained by the juxtaposition of four copies of $K(t)$, which are
labeled as $K_{t}^{1}$, $K_{t}^{2}$, $K_{t}^{3}$, and $K_{t}^{4}$,
respectively. Then we can write the sum $D_{t+1}$ as
\begin{equation}\label{eq:app3}
  D_{t+1} = 4\,D_t + \Delta_t\,,
\end{equation}
where $\Delta_t$ is the sum over all shortest paths whose end points
are not in the same $K_{t}$ branch. The solution of
Eq.~\eqref{eq:app3} is
\begin{equation}\label{eq:app4}
  D_t = 4^{t-1} D_1 + \sum_{x=1}^{t-1} 4^{t-x-1} \Delta_x\,.
\end{equation}
The paths that contribute to $\Delta_t$ must all go through at least
one of the three edge nodes (i.e., $X$, $Y$ and $Z$ in
Fig.~\ref{copy}) at which the different $K_t$ branches are
connected. The analytical expression for $\Delta_t$, called the
length of crossing paths, is found below.

Denote $\Delta_t^{\alpha,\beta}$ as the sum of length for all
shortest paths with end points in $K_t^{\alpha}$ and $K_t^{\beta}$,
respectively. If $K_t^{\alpha}$ and $K_t^{\beta}$ meet at an edge
node, $\Delta_t^{\alpha,\beta}$ rules out the paths where either end
point is that shared edge node. For example, each path contributing
to $\Delta_t^{1,2}$ should not end at node $X$. If $K_t^{\alpha}$
and $K_t^{\beta}$ do not meet, $\Delta_t^{\alpha,\beta}$ excludes
the paths where either end point is any edge node. For instance,
each path contributive to $\Delta_t^{2,3}$ should not end at node
$X$ or $Y$. Then the total sum $\Delta_t$ is
\begin{align}
\Delta_t =& \,\Delta_t^{1,2} + \Delta_t^{1,3} + \Delta_t^{1,4}+
\Delta_t^{2,3} + \Delta_t^{2,4}+\Delta_t^{3,4}. \label{eq:app5}
\end{align}

By symmetry, $\Delta_t^{1,2} = \Delta_t^{1,3} = \Delta_t^{1,4}$ and
$\Delta_t^{2,3} = \Delta_t^{2,4}=\Delta_t^{3,4}$, so that
\begin{equation}\label{eq:app6}
\Delta_t = 3 \Delta_t^{1,2} + 3 \Delta_t^{2,3}\,.
\end{equation}
In order to find $\Delta_t^{1,2}$ and $\Delta_t^{2,3}$, we define
\begin{align}
s_t = \sum_{\stackrel{i \in K(t)}{i\ne X}}d_{iX}\,. \label{eq:app7}
\end{align}
Considering the self-similar network structure, we can easily know
that at time $t+1$, the quantity $s_{t+1}$ evolves recursively as
\begin{eqnarray}
s_{t+1}
&=&2\,s_t+\left[s_t+(N_t-1)\right]+\left[s_t+(N_t-1)\right]\nonumber\\
&=&4\,s_t+4\times 4^{t}.\label{eq:app8}
\end{eqnarray}
Using $s_1=12$, we have
\begin{eqnarray}
s_t=4^{t}\,(t+2).
\end{eqnarray}
On the other hand, by definition given above, we have
\begin{eqnarray}
  \Delta_t^{1,2} &=& \sum_{\substack{i \in K_t^{1},\,\,j\in
      K_t^{2}\\ i,j \ne X}} d_{ij}\nonumber\\
  &=& \sum_{\substack{i \in K_t^{1},\,\,j\in
      K_t^{2}\\ i,j \ne X}} (d_{iX} + d_{jX}) \nonumber\\
  &=& (N_t-1)\sum_{\substack{i \in K_t^{1}\\ i \ne X}} d_{iX} + (N_t-1) \sum_{\substack{j \in K_t^{2}\\ j \ne X}} d_{jX} \nonumber\\
  &=& 2(N_t-1)\sum_{\substack{i \in K_t^{1}\\ i \ne X}} d_{iX}\nonumber\\
  &=& 2(N_t-1)\,s_t,
\label{eq:app9}
\end{eqnarray}
and
\begin{eqnarray}
  \Delta_t^{2,3} &=& \sum_{\substack{i \in K_t^{2},\,i \ne X\\ j\in
      K_t^{3},\,j \ne Y}} d_{ij}\nonumber\\
  &=& \sum_{\substack{i \in K_t^{2},\,i \ne X\\ j\in
      K_t^{3},\,j \ne Y}} (d_{iX} + d_{XY}+ d_{jY}) \nonumber\\
  &=& 2(N_t-1)\,s_t+(N_t-1)^2\,,
\label{eq:app10}
\end{eqnarray}
where $d_{XY}=1$ has been used. Substituting Eqs.~(\ref{eq:app9})
and (\ref{eq:app10}) into Eq.~(\ref{eq:app6}), we obtain
\begin{eqnarray}\label{eq:app12}
\Delta_t &=& 12(N_t-1)\,s_t+3\,(N_t-1)^2\,\nonumber\\
&=&(12t+27)\times16^{t}.
\end{eqnarray}
Inserting Eq.~(\ref{eq:app12}) for $\Delta_x$ into
Eq.~(\ref{eq:app4}) and using $D_1 =72$, we have
\begin{equation}
 D_t = \frac{4^t}{3}\times \left(2+7\times4^t+ 6t\times4^t\right). \label{eq:app13}
\end{equation}
Inserting Eq.~\eqref{eq:app13} into Eq.~\eqref{eq:app1}, one can
obtain the analytical expression for $d_t$ as shown in
Eq.~\eqref{APL}.

\end{document}